\documentclass[twocolumn]{article}
\RequirePackage{xcolor}
\usepackage[a4paper, total={18cm, 26cm}]{geometry}
\usepackage{amsmath,amsfonts}
\usepackage{algorithmic}
\usepackage{algorithm}
\usepackage{array}
\usepackage[caption=false,font=normalsize,labelfont=sf,textfont=sf]{subfig}
\usepackage{textcomp}
\usepackage{stfloats}
\usepackage[exponent-product=\cdot]{siunitx}
\usepackage{url}
\usepackage{verbatim}
\usepackage{graphicx}
\usepackage{hyperref}
\hypersetup{colorlinks=false,pdfborder={0 0 0}} 
\usepackage{breqn,multirow}
\usepackage{circuitikz}
\usepackage{tikz}
\usetikzlibrary{patterns}
\usepackage{pgfplots}
\pgfplotsset{compat=1.17}
\usepackage{tudacolors}
\usepackage{tuda-pgfplots}
\usepackage{graphicx}
\usepackage[style=ieee,url=false,doi=false,eprint=false]{biblatex}
\DeclareSourcemap{
  \maps[datatype=bibtex]{
	\map[overwrite]{
	  \step[fieldsource=shortjournal, fieldtarget=journaltitle]
	}
  }
}
\AtEveryBibitem{\clearfield{day}}
\AtEveryBibitem{\clearfield{month}}
\AtEveryBibitem{\clearfield{issn}}
\AtEveryBibitem{\clearfield{series}}
\AtEveryBibitem{\clearfield{isbn}}
\AtEveryBibitem{\clearfield{organization}}
\AtEveryBibitem{\clearfield{institution}}

\usepackage{eso-pic}
\AddToShipoutPicture*{\footnotesize\sffamily\raisebox{28.5cm}{\hspace{1.3cm}\fbox{\parbox{\textwidth}{\copyright 2023 IEEE.  Personal use of this material is permitted.  Permission from IEEE must be obtained for all other uses, in any current or future media, including reprinting/republishing this material for advertising or promotional purposes, creating new collective works, for resale or redistribution to servers or lists, or reuse of any copyrighted component of this work in other works.}}}}
\begin{document}

\title{\textbf{Low-Frequency Stabilization of Dielectric Simulation Problems with Conductors and Insulators}}
\author{Devin~Balian, Melina~Merkel, Jörg~Ostrowski, Herbert~De~Gersem and Sebastian~Schöps
\thanks{The first, second, fourth and fifth author are with the Department of Electrical Engineering and Information Technology at the Technical University of Darmstadt, Germany. The third author is with Siemens Digital Industries Software, Switzerland (contact e-mail: sebastian.schoeps@tu-darmstadt.de).}%
}

\maketitle

\begin{abstract}
When simulating resistive-capacitive circuits or electroquasistatic problems where conductors and insulators coexist, one observes that large time steps or low frequencies lead to numerical instabilities, which are related to the condition number of the system matrix. Here, we propose several stable formulations by scaling the equation systems. This enables a reliable calculation of solutions for very low frequencies (even for the static case), or large time steps. Numerical experiments underline the findings. 
\end{abstract}

\bigskip


\section{Introduction}
Low-frequency electroquasistatic field or resistive-capacitive circuit simulations, for example of high-voltage applications, are well established in academia and industry, \cite{Haus_1989aa,Dirks_1996aa,van-Rienen_1996aa,Clemens_1998aa,Monga_2006aa,Weida_2009aa,Christen_2010aa,Zhang_2010aa}. 
The underlying field approximation disregards inductive effects and therefore allows a simplified formulation based only on the scalar-valued electric potential. However, if coupled capacitive, inductive and resistive phenomena are relevant, classical Maxwell formulations are necessary or -- if wave propagation is negligible -- mixed formulations of Darwin-type that combine the electro- and magnetoquastistatic cases, see, e.g., \cite{Clemens_2022aa} and the references therein. For full-wave formulations in frequency domain, it is well-known that they exhibit a low-frequency instability. The issue follows from the fact that Maxwell equations decouple in the static limit into three separate magnetostatic, electrostatic and stationary current problems. In particular, the magnetostatic problem requires gauging which is well understood in the limit case, but is (numerically) cumbersome for very small but non-zero frequencies. Several stabilized formulations have been proposed, for example by Hiptmair \cite{Hiptmair_2008aa}, Jochum \cite{Jochum_2015aa}, Eller \cite{Eller_2017aa} which was later also used by Stysch \cite{Stysch_2022aa} and Zhao \cite{Zhao_2019aa}.

\begin{figure}
	\centering
	\begin{circuitikz}
		\draw (0,0)
			to[I,l=$I$,-] (0,1.8)
			to[short,-*] (1.8,1.8)
			node[label={[font=\footnotesize]above:1}] {}
			to[C,l=$C_1$,-*] (2*1.8,1.8)
			node[label={[font=\footnotesize]above:2}] {}
			to[C,l=$C_2$,-] (2*1.8,0)
			to[short,-*] (1.8,0)
			to[short,-] (0,0);
		\draw (1.8,1.8)
			to[R,l=$R_3$] (1.8,0)
			node[ground]{};
	\end{circuitikz}
	\caption{A simple RC circuit with problematic low-frequency behavior.}
	\label{fig:rc}
\end{figure}
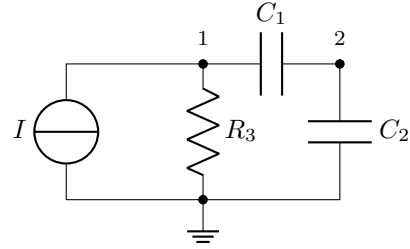

This paper investigates a similar low-frequency instability for electroquasistatic field and circuit formulations that is not related to gauging. The problem was initially observed in \cite{Ostrowski_2021aa,Kasolis_2021aa}: in the static limit the electroquasistatic field problem turns into a stationary current problem but only within the conductors. One loses control over the fields in insulating domains since the displacement current density vanishes in the static limit. This gives rise to stability issues as in the full-wave case. In that regime, iterative solvers may suffer from excessive floating-point rounding-off errors \cite{Saad_2000aa}. The solution then exhibits visible defects (as, e.g., in \autoref{fig:roth2_bushing}(c)), or may feature hidden defects, which are even more dangerous, because they may trigger wrong engineering decisions. 
In \cite{Ostrowski_2021aa} the low-frequency breakdown was mitigated by additional constraints enforcing charge-neutrality in the non-conducting domain, however, at the cost of additional unknowns. 
In this paper, we propose several scalings of the equations involving the frequency (or time step size) which do not introduce additional unknowns and are easy to implement. Eventually we demonstrate the effectiveness of this new approach.

Let us motivate the problem using the simple current-driven circuit from \autoref{fig:rc} in frequency domain. We do not use the law for capacitors in series to simplify the model. 
Then, nodal analysis yields
\begin{align}
	\left(
	\begin{bmatrix}
		R_3^{-1} & 0\\
		0 & 0
	\end{bmatrix}
	+
	j\omega \begin{bmatrix}
		C_1 & -C_1\\
		-C_1 & C_1+C_2
	\end{bmatrix}
	\right)
	\begin{bmatrix}
		\phi_1\\
		\phi_2
	\end{bmatrix}
	=   
	\begin{bmatrix}
		-I\\
		0
	\end{bmatrix}
	\label{eq:rc_circuit}
\end{align}
where $\phi_n$ is the electric potential related to node $n$, $i_{\mathrm{src}}$ is a given current with frequency $\omega$ and $j=\sqrt{-1}$ is the imaginary unit; $R_k$ and $C_k$ are the resistance and capacitances of branch $k$, respectively. The resulting system is uniquely solvable for any frequency $\omega>0$ but the condition number of the linear equation system explodes for $\omega\to0$, e.g.,
$$
\kappa(\omega)=1+\frac{1}{2\omega RC}+\mathcal{O}(\omega) 
$$
for $C_1=C_2=C$, $R_3=R$ and $R^{-1}>2\omega C$ and using the $1$-norm. In the limit, the condition number is $\kappa(0)=\infty$. This reflects the fact that we can only recover Ohm's law $\phi_1=R_3i$ from \eqref{eq:rc_circuit} but there is no equation for $\phi_2$, i.e., the variable is undefined. Luckily, the problem can be fixed almost trivially when scaling the equation system appropriately, e.g. by Jacobi-type preconditioners.

The paper is structured as follows: we recall in the next section common resistive-capacitive circuit and electroquasistatic field formulations and introduce a partitioning according to their insulating and conductive parts. Several options for stabilization are introduced in Section III and are applied to examples in Section IV. Finally, Section V concludes this paper.

\section{Problem Formulation}
In this section we recapitulate formulations of resistive-capacitive (RC) circuits and electroquasistatic (EQS) fields. 
\subsection{RC Circuits}
We start with circuits in frequency domain described by modified nodal analysis which is the most common formalism used in academic and industrial SPICE-like solvers, \cite{Ho_1975aa}. The problem is: find $\boldsymbol{\varphi}\in\mathbb{C}^{N_{\varphi}},\mathbf{i}_{\mathrm{V}}\in\mathbb{C}^{N_{\mathrm{V}}}$ such that
\begin{subequations}
\begin{align}
	\label{eq:mna1}
	\left(
	\mathbf{A}_{\mathrm{R}}
	\mathbf{G}
	\mathbf{A}_{\mathrm{R}}^{\top}
	\right)
	\boldsymbol{\varphi}
	+
	j\omega\mathbf{A}_{\mathrm{C}}
	\mathbf{C}
	\mathbf{A}_{\mathrm{C}}^{\top}
	+
	\mathbf{A}_{\mathrm{V}}
	\mathbf{i}_{\mathrm{V}}
	&=
	- \mathbf{A}_{\mathrm{I}}
	\mathbf{i}_{\mathrm{src}}
	\\
	\mathbf{A}_{\mathrm{V}}^\top
	\boldsymbol{\varphi}
	&=
	\mathbf{v}_{\mathrm{src}}
	\label{eq:mna2} 
\end{align}
\end{subequations}
with incidence matrices $\mathbf{A}_k$ for element type $k\in\{\mathrm{R},\mathrm{C},\mathrm{I},\mathrm{V}\}$, the $N_{\varphi}$-dimensional vector of unknown nodal potentials $\boldsymbol{\varphi}$, the $N_{\mathrm{V}}$-dimensional vector of unknown currents through voltage sources $\mathbf{i}_{\mathrm{V}}$, given (complex) currents and voltage $\mathbf{i}_{\mathrm{src}}$ and $\mathbf{v}_{\mathrm{src}}$ due to sources, and diagonal matrices of conductances  $\mathbf{G}$ and capacitances $\mathbf{C}$. 
Note that such a circuit must fulfill several compatibility conditions, e.g., currents may not be prescribed on capacitive branches if $\omega\to0$. 

\subsection{Electroquasistatic Fields}
The field equivalent of a RC circuit is the electroquasistatic field formulation, see, e.g., \cite{Haus_1989aa,Dirks_1996aa}, given by
\begin{subequations}
\begin{align}
	\label{eq:eqs_mono1}
	-\nabla\cdot\Bigl(\sigma\nabla\varphi\Bigr) -j\omega\nabla\cdot\Bigl(\varepsilon\nabla\varphi\Bigr) &= \nabla\cdot\mathbf{J}_{\textrm{src}} && \text{on }\Omega
	\\
	\label{eq:eqs_mono2}
	\varphi&=0 && \text{on }\partial\Omega
\end{align}
\end{subequations}
where $\varphi$ is the electric scalar potential, $\mathbf{J}_{\textrm{src}}$ the source current density, $\varepsilon>0$
and $\sigma\geq0$ are the space-dependent material coefficients for permittivity and conductivity, respectively. Again for simplicity of notation, we have equipped the problems with a homogeneous Dirichlet condition and we do not consider electrodes at constant potential, which necessitate floating-potential conditions. The introduction thereof will be sketched below.

\begin{figure}
	\centering
	\scalebox{0.9}{\begin{tikzpicture}
		\node at (5,4.1) {$\partial\Omega$};
		\draw[clip, preaction={draw}] plot[smooth cycle, tension=2] coordinates {(1,2) (3,4.5) (6,1)};
		\draw [fill=black!20] plot [smooth cycle, tension=2] coordinates {(-1.0,2.2) (3,3.6) (3,1)};
		\node[align=center] at (4.5,1) {$\Omega_{2}$\\$\varepsilon_{2}=\textrm{const.}$\\$\sigma_{2}=0$};
		\node[align=center] at (2.5,2.3) {$\Omega_{1}$\\$\varepsilon_{1}=\textrm{const.}$\\$\sigma_{1}>0$};
		\node at (3.3,3.7) {$\Gamma_{12}$};
	\end{tikzpicture}}
	\vspace{-1.5em}
	\caption{Abstract representation of the computational domain.}
	\label{fig:domain}
\end{figure}
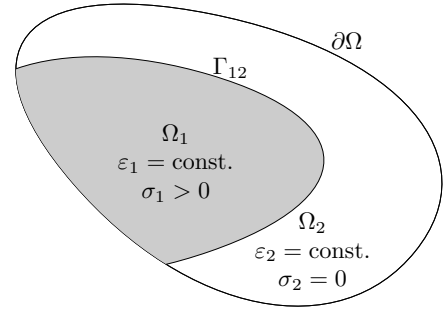

Let us investigate the case in which the computational domain $\bar\Omega=\bar\Omega_1\cup\bar\Omega_2$ contains a homogeneous conductor $\Omega_1$ and an insulator $\Omega_2$, see \autoref{fig:domain}. The problem reads
\begin{subequations}
\begin{align}
	\label{eq:eqs_sep1}
	-\nabla\!\cdot\!\Bigl(\sigma_1\nabla\varphi_1\Bigr) -j\omega\nabla\!\cdot\!\Bigl(\varepsilon_1\nabla
	\varphi_1\Bigr) &=\nabla\!\cdot\!\mathbf{J}_{\textrm{src},1} \!\!\!\!&& \text{on }\Omega_1
	\\
	\label{eq:eqs_sep2}
	-j\omega\nabla\!\cdot\!\Bigl(\varepsilon_2\nabla
	\varphi_2\Bigr) &=\nabla\!\cdot\!\mathbf{J}_{\textrm{src},2} \!\!\!\!&& \text{on }\Omega_2
	\\
	\varphi_1-\varphi_2&=0 && \text{on }\Gamma_{12}
	\\
	\mathbf{n}\!\cdot\!(\mathbf{J}_{\textrm{tot},1}-\mathbf{J}_{\textrm{tot},2})&=0 && \text{on }\Gamma_{12}
	\\
	\label{eq:eqs_sep5}
	\varphi_{1}=\varphi_{2}&=0 && \text{on }\partial\Omega
\end{align}
\end{subequations}
where $\varphi_p$, $\mathbf{J}_{\textrm{src},p}$, $\varepsilon_p$ and $\sigma_p$ are defined on each domain $\Omega_p$ separately. They are glued by interface conditions on $\Gamma_{12}=\bar\Omega_1\cap\bar\Omega_2$ using the total current density
$\mathbf{J}_{\textrm{tot},p}=-(\sigma_p+j\omega\varepsilon_p)\nabla\varphi_p$ and the interface's normal vector $\mathbf{n}$. In this formulation, it becomes apparent that we must require a compatibility condition $\nabla\cdot\mathbf{J}_{\textrm{src},2}=0$ in the limit $\omega\to0$; such a property is called `divergence-free in stationary limit' in \cite{Eller_2017aa}.

There are two options for discretization: (a) we derive a monolithic weak formulation for \eqref{eq:eqs_mono1} with basis functions on the total domain $\Omega$ and allow jumping material coefficients or (b) we apply finite elements (FE) to both equations in \eqref{eq:eqs_sep1}-\eqref{eq:eqs_sep5} separately by using basis functions with support in $\Omega_1$ and $\Omega_2$ and glue them afterwards weakly. We will discuss only the first option (a) since this is implemented in most academic or industrial EQS solvers. The regularizations proposed in Section~\ref{sec:stabil} will work in both cases. 

The weak formulation of the problems reads: find $\varphi \in H^1_0(\Omega)$ such that
\begin{align}
	\label{eq:fem}
	\!\int_{\Omega}\!\!\nabla \varphi' \cdot \sigma\nabla\varphi
	+ 
	j\omega
	\nabla \varphi' \cdot \varepsilon\nabla\varphi\;\mathrm{d}\Omega
	&=
	-\!\!\int_{\Omega}\!\!\nabla\varphi' \cdot \mathbf{J}_{\textrm{src}}\;\mathrm{d}\Omega
\end{align}
for all $\varphi'\in H^1_0(\Omega)$, i.e., functions from the space of square-integrable functions with square-integrable gradient that vanish on the boundary \cite{Monk_2003aa}. Restricting ourselves to an $N$-dimensional subspace $V\subset H^1_0(\Omega)$, we can rewrite this system in matrix/vector notation, i.e., find $\boldsymbol{\varphi}$
such that
\begin{align}
	\label{eq:eqs_fem}
	\left(\mathbf{K}
	+
	j\omega\mathbf{M}\right)
	\boldsymbol{\varphi}
	=
	\mathbf{r}
\end{align}
with the usual FE matrices, given by the entries
\begin{subequations}
\begin{align}
	K_{mn} &=\int_\Omega\!\!\nabla v_m \cdot \sigma\nabla v_n\,\mathrm{d}\Omega,
	\;\\
	M_{mn} &=\int_\Omega\!\!\nabla v_m \cdot \varepsilon\nabla v_n\,\mathrm{d}\Omega
\end{align}
\end{subequations}
and the right-hand-side
\begin{align}
	r_{m}=-\int_\Omega\!\!\nabla v_m \cdot \mathbf{J}_{\textrm{src}}\,\mathrm{d}\Omega
\end{align}
with nodal basis functions $v_n\in V$ and $m,n=1,\ldots,N$.

In the electrostatic limit case, conductors at a constant but unknown potential are modeled by floating-potential conditions. In the electroquasistatic case, it makes sense to model highly conductive parts by floating-potential conditions as well. To accommodate for them, as well as for inhomogeneous Dirichlet conditions, the model needs small adaptations, which are sketched here but not spelled out in the notations used further below. The floating-point conditions are introduced by considering a subspace of $H_0^1(\Omega)$ where the potentials are constant at the floating-potential parts. Inhomogeneous Dirichlet conditions are considered by shifting the trial space \cite{Brenner_2008aa}. In the discrete setting, floating-potential conditions are applied to $V$ by gluing together all nodal basis functions coinciding with the same electrode, thereby reducing the number of degrees of freedom \cite{Dular_1998aa}. This technique can be efficiently implemented by a few algebraic manipulations to the system matrix \cite{De-Gersem_2003ac}.

\subsection{Decomposition}
By renumbering of degrees of freedom, both systems \eqref{eq:mna1}-\eqref{eq:mna2} and \eqref{eq:eqs_fem} allow a block-matrix partitioning in the form
\begin{align}
	\left(
		\begin{bmatrix}
			\mathbf{K}_{11} & \mathbf{0}\\
			\mathbf{0} & \mathbf{0}\\
		\end{bmatrix}
		+
		j\omega\begin{bmatrix}
			\mathbf{M}_{11} & \mathbf{M}_{12}\\
			\mathbf{M}_{12}^{\top} & \mathbf{M}_{22}\\
		\end{bmatrix}
	\right)
	\begin{bmatrix}
		\boldsymbol{\varphi}_{1}\\
		\boldsymbol{\varphi}_{2}
	\end{bmatrix}  
	=
    \begin{bmatrix}
		\mathbf{r}_{1}\\
		\mathbf{r}_{2}
	\end{bmatrix}.
	\label{eq:partition}
\end{align}
The zero contributions in the conductivity-related operator are a consequence of the insulator in $\Omega_2$.
In practice, an explicit reordering is not necessary. One may work with index sets. For example in the case of low-order FE, we define the index sets of degrees of freedom in $\Omega_2$ and $\Omega_1$ as
\begin{subequations}
\begin{align}
	\mathcal{I}_2&=\{1\leq n\leq N\;|\;\mathrm{supp}(v_n)\in\Omega_2, v_n\in V\}\\
	\mathcal{I}_1&=\{1\leq n\leq N\;|\;n\not\in \mathcal{I}_2\}
\end{align}
\end{subequations}
and thus obtain the structure of system \eqref{eq:eqs_fem} implicitly. Note, those index sets put the degrees of freedom on the interface $\Gamma_{12}$ in $\boldsymbol{\varphi}_{1}$ since the basis functions related to the interface are not fully contained on $\Omega_2$. This is necessary to obtain the zero blocks in \eqref{eq:partition}.

Finally, we observe that \eqref{eq:partition} has the same structure as \eqref{eq:rc_circuit} and thus the same issue for $\omega\to0$ must be expected. Indeed, the bad condition number for low frequencies was observed in \cite{Ostrowski_2021aa}. This problem will be mitigated in the next section.

\section{Low-Frequency Stabilization}\label{sec:stabil}
While \cite{Kasolis_2021aa} proposes a two step approach, we keep the original system but multiply equations and unknowns including powers of $\omega$ such that no equation vanishes for $\omega\to0$. We investigate in the first section scalar-valued (`scaling') and in the second section matrix-valued multiplications (`preconditioning'). 
Let us start with four scalars $a_1, a_2, b_1, b_2\in\mathbb{C}$ and multiply \eqref{eq:partition} as follows
\begin{align}
	\!\left(
		\begin{bmatrix}
			a_1b_1\mathbf{K}_{11} & \!\!\!\mathbf{0}\\
			\mathbf{0} & \!\!\!\mathbf{0}\\
		\end{bmatrix}
        \!
		+
		j\omega\!
        \begin{bmatrix}
			a_1b_1\mathbf{M}_{11} & a_1b_2\mathbf{M}_{12}\\
			a_2b_1\mathbf{M}_{12}^{\top} & a_2b_2\mathbf{M}_{22}\\
		\end{bmatrix}
	\right)\!\!
	\begin{bmatrix}
		\boldsymbol{\xi}_{1}\\
		\boldsymbol{\xi}_{2}
	\end{bmatrix}  
	=
	\begin{bmatrix}
		a_1\mathbf{r}_{1}\\
		a_2\mathbf{r}_{2}\\
	\end{bmatrix}
	\nonumber
\end{align}
where the new (scaled) unknowns are $\boldsymbol{\xi}_k=b_k^{-1}\boldsymbol{\varphi}_k$. We propose the following scalings in frequency domain:
\begin{itemize}
    \item[(i)] symmetric: $a_2=b_2=\omega^{-1/2}$
    and $a_1=b_1$=1
    \item[(ii)] non-symmetric: $a_2=\omega^{-1}$
    and $a_1=b_1=b_2=1$
\end{itemize}
{which must be applied on the analytical level, i.e., the products of powers of $\omega$ must be determined before matrix assembly to avoid numerical errors or division by zero.} Then, both variants ensure that no equation is lost in the limit $\omega\to0$ since it always holds $\omega a_2b_2>0$. Additionally, the first approach (i) maintains the symmetry of the resulting linear equation system. This allows to save memory and enables the application of dedicated solution methods, e.g., Cholesky factorization or preconditioned conjugate gradients. It is also in good agreement with the ideas applied in the full-wave case, for example, a scaling by fractional powers of $\omega$ was similarly applied in \cite{Eller_2017aa}. However, a reconstruction of the potential in the limit $\omega=0$ is not possible. This is a natural consequence of the fact that it is not well-defined from the start. On the other hand, one may be interested in obtaining the electrostatic field solution in $\Omega_2$ for $\omega=0$. This is guaranteed by variant (ii) which keeps the original unknowns at the price of losing symmetry. 

Note that the same idea can also be applied in time domain simulation, where similar stability issues may occur for very large time step sizes. Let us stress that this is rather related to the formulation than to the numerical solution method. More precisely, it is not related to the well-known time step restriction of many (explicit) time-stepping methods applied to stiff differential equations \cite{Hairer_1996aa}. 
In the time discrete setting, one scales the system matrix in every time step by powers of the step size instead of the frequency. For example, one time step of the implicit Euler method scaled according to (i) reads
\begin{align}
    \nonumber
	\begin{bmatrix}
		\mathbf{K}_{11}+\frac{1}{\delta t_{l}}\mathbf{M}_{11} &\frac{1}{\sqrt{\delta t_l}}\mathbf{M}_{12}\\
		\frac{1}{\sqrt{\delta t_l}}\mathbf{M}_{12}^{\top} & \mathbf{M}_{22}\\
	\end{bmatrix}
	\begin{bmatrix}
		\boldsymbol{\varphi}_{1}(t_{l+1})\\
		\boldsymbol{\xi}_{2}(t_{l+1})
	\end{bmatrix}
    \\
	\!=
	\begin{bmatrix}
		\frac{1}{\delta t_{l}}\mathbf{M}_{11} & \frac{1}{\sqrt{\delta t_l}}\mathbf{M}_{12}\\
		\frac{1}{\sqrt{\delta t_l}}\mathbf{M}_{12}^{\top} & \mathbf{M}_{22}\\
	\end{bmatrix}
  	\begin{bmatrix}
		\boldsymbol{\varphi}_{1}(t_{l})\\
		\boldsymbol{\xi}_{2}(t_{l})
	\end{bmatrix}
	+
    \begin{bmatrix}
		\mathbf{r}_{1}(t_{l+1})\\
		\sqrt{\delta t_l}\mathbf{r}_{2}(t_{l+1})\\
	\end{bmatrix}
    \label{eq:euler}
\end{align}
with time step size $\delta t_{l}=t_{l+1}-t_{l}$ and $\boldsymbol{\xi}_2=\frac{1}{\sqrt{\delta t_l}}\boldsymbol{\varphi}_2$. 

While not strictly related to the low-frequency breakdown, involving material constants in the scaling helps to further equilibrate the spectrum of the system matrix and thus to reduce the condition number. For example, the following scalar choices are obvious candidates for the FE problem:
\begin{itemize}\itemsep0.3em
    \item[(iii)] symmetric with material: $a_1=b_1=(\sigma_1{+j\omega\varepsilon_1})^{-1/2}$ and $a_2=b_2=(\varepsilon_2 j\omega)^{-1/2}$
    \item[(iv)] non-symmetric with material: $a_1=(\sigma_1{+j\omega\varepsilon_1})^{-1}$, $a_2=(\varepsilon_2 j\omega)^{-1}$ and $b_1=b_2=1$
\end{itemize}  
Variant (iv) is closely related to Jacobi-type preconditioning but applied here \textit{before} the matrix assembly \cite[Section 4.1]{Saad_2000aa}. Jacobi-type preconditioning uses the inverse of the diagonal entries, e.g.,
\begin{align}
    \label{eq:jacobi1}
    a_1&=\Bigl(\mathrm{diag}(\mathbf{K}_{11})+j\omega\,\mathrm{diag}(\mathbf{M}_{11})\Bigr)^{-1},
    \\
    a_2&=\frac{1}{j\omega}\Bigl(\mathrm{diag}(\mathbf{M}_{22})\Bigr)^{-1}
    \label{eq:jacobi2}
\end{align}
for left multiplication ($b_1=b_2=1$) or the square roots to multiply from left and right in the spirit of variant (iii). If iterative methods shall be used to solve the resulting equation system, then this idea can be taken even further. One may use incomplete inverses of matrix blocks as (left) preconditioners \cite[Section 10.3]{Saad_2000aa}, 
 e.g.,
\begin{itemize}\itemsep0.3em
    \item[(v)] frequency dependent block preconditioner:\\ $a_1=(\mathbf{K}_{11}+j\omega\mathbf{M}_{11})^{-1}$, $a_2=\frac{1}{j\omega}\mathbf{M}_{22}^{-1}$ and $b_1=b_2=1$ 
    \item[(vi)] frequency independent block preconditioner:\\ $a_1=(\mathbf{K}_{11}+j\omega_0\mathbf{M}_{11})^{-1}$ with a fixed $\omega_0$, $a_2=\frac{1}{j\omega}\mathbf{M}_{22}^{-1}$ and $b_1=b_2=1$
\end{itemize}
where we have assumed that such (incomplete) block inverses can be computed. This is for example the case when all conducting parts touch a Dirichlet boundary.

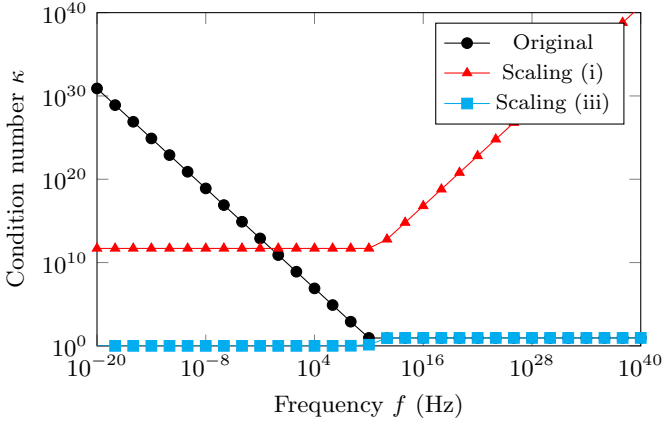
\begin{figure}
	\centering
	\begin{tikzpicture}
	\begin{loglogaxis}[
            width=0.99\columnwidth,
            height=6cm,
            xlabel={Frequency $f$ (Hz)},
            ylabel={Condition number $\kappa$}, 
            xmin=1e-20, xmax=1e40,
            ymax=1e40, ymin=1,
            ytick={1e0, 1e10, 1e20, 1e30, 1e40},
            tick label style={font=\small},
            legend pos={north east},
            legend style={font=\footnotesize}, 
            label style={font=\small},
            every x tick scale label/.style={at={(1,0)},
            anchor=north,yshift=-5pt,inner sep=0pt}]
			\addplot [black, mark=*] table [x index=0, y index=2, col sep=comma, header=true]{data/condition_circuit.csv};
			\addplot [red, mark=triangle*] table [x index=0, y index=3, col sep=comma, header=true]{data/condition_circuit.csv};
			\addplot [cyan, mark=square*] table [x index=0, y index=4, col sep=comma, header=true]{data/condition_circuit.csv};
			\legend{
				Original\\ 
				Scaling (i) \\
				Scaling (iii) \\
			}
	\end{loglogaxis}
	\end{tikzpicture}
	\caption{Condition number of the RC circuit example ($C=\SI{1e12}{\farad}$ and $R=\SI{1}{\ohm}$) in dependence of frequency.}
	\label{fig:condition_circuit}
\end{figure}

An advantage of preconditioner (vi) is that the computational cost of repeated factorization in frequency sweeps can be significantly reduced since there is no dependence on $\omega$ within the inverse. Furthermore, when choosing $\omega_0=0$, the factorization can be carried out in non-complex arithmetics. On the other hand, one must expect that the larger the distance $|\omega-\omega_0|$, the worse the performance. 

Note that we have assumed the domain to consist of one conducting and one non-conducting subdomain. However, stabilization variants (i-iv) can easily be generalized to multiple domains with different material properties; variants (v-vi) consider this automatically.

\begin{figure}
    \centering
    \begin{tikzpicture}[scale=0.13]
    \begin{scope}
        \draw (0,0) rectangle (22,22);
        \fill[TUDa-9b, opacity=0.5] (0,10) rectangle (22, 12);
        \fill[TUDa-1b, opacity=0.5] (10,0) rectangle (12, 22);
        \draw[TUDa-1b, very thick] (10,0) -- (10,22);
        \draw[TUDa-1b, very thick] (12,0) -- (12,22);       
        \draw[red, very thick] (0,0) -- (0,22);
        \draw[red, very thick] (22,0) -- (22,22);
        \pattern[pattern=north east lines, pattern color=red] 
            (0,0) rectangle (-1.5,22);
        \pattern[pattern=north east lines, pattern color=red] 
            (22,0) rectangle (23.5,22);
        \draw[xshift=-14em, align = right] 
            (0,11) node{$\varphi_{\mathrm{bc},1}$};
        \draw[xshift=14em, align = left] 
            (22,11) node{$\varphi_{\mathrm{bc},2}$};            
        \draw[thick, ->] (24,21) -- (11,15)
            node[at start, xshift=1em]{$\varepsilon_{\mathrm{i}}$};
        \draw[thick, ->] (24,17) -- (15,15)
            node[at start, xshift=1em]{$\varepsilon_{\mathrm{o}}$};
        \draw[thick, ->] (24,5) -- (15,11)
            node[at start, xshift=1em]{$\sigma_{\mathrm{o}}$};
        \draw[thick, ->] (24,1) -- (11,11)
            node[at start, xshift=1em]{$\sigma_{\mathrm{i}}$};       
        \draw [
            decorate,
            decoration={brace,amplitude=5pt,mirror,raise=5pt}
        ] (0,0) -- (9.5,0) node[midway,yshift=-2em]{$d_{\mathrm{o}}$};
        \draw [
            decorate,
            decoration={brace,amplitude=5pt,mirror,raise=5pt}
        ] (10,0) -- (12,0) node[midway,yshift=-2em]{$d_{\mathrm{i}}$};
        \draw [
            decorate,
            decoration={brace,amplitude=5pt,mirror,raise=5pt}
        ] (12.5,0) -- (22,0) node[midway,yshift=-2em]{$d_{\mathrm{o}}$};  
        \draw[thick, ->, xshift=-5cm, yshift=-5cm] (0,0) -- (5,0)
            node[at end,xshift=5pt]{$x$};
        \draw[thick, ->, xshift=-5cm, yshift=-5cm] (0,0) -- (0,5)
            node[at end,yshift=5pt]{$y$};        
    \end{scope}
    
    \begin{scope}[xshift=35cm]
        \draw[fill=TUDa-1b, opacity=0.5] (0,0) rectangle (22,22);
        \fill[TUDa-9b, opacity=0.5] (10,10) rectangle (12, 12);
        \draw [
            decorate,
            decoration={brace,amplitude=5pt,mirror,raise=5pt}
        ] (10,10) -- (12,10) node[midway,yshift=-2em]{$d_{\mathrm{i}}$};
        \draw [
            decorate,
            decoration={brace,amplitude=5pt,mirror,raise=5pt}
        ] (0,0) -- (22,0) node[midway,yshift=-2em]{$d$};        
        \draw[thick, ->, xshift=-5cm, yshift=-5cm] (0,0) -- (5,0)
            node[at end,xshift=5pt]{$z$};
        \draw[thick, ->, xshift=-5cm, yshift=-5cm] (0,0) -- (0,5)
            node[at end,yshift=5pt]{$y$};       
    \end{scope}
    \end{tikzpicture}
    \vspace*{-0.25em}
    \caption{Two-dimensional representation of the test configuration. Shown is the side view (left) and the front view (right). This setup is based on \cite{Ostrowski_2021aa}.}
    \label{fig:anordnung}
\end{figure}
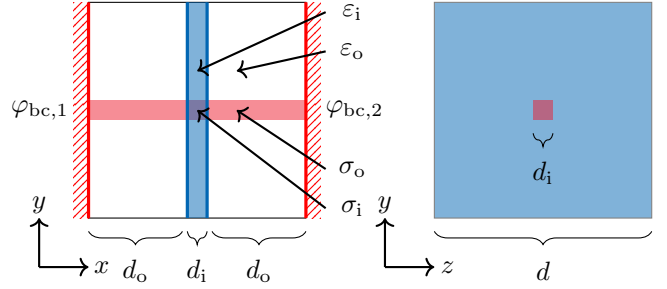

\section{Examples}
We discuss the cases of an RC circuit and two electroquasistatic field problems: an academic toy example and a high-voltage bushing.

\subsection{RC Circuit}
Let us start with the simple circuit example \eqref{eq:rc_circuit} in frequency domain using the simplification $C_1=C_2=C$ and $R_3=R$. The incidence matrices corresponding to \eqref{eq:mna1}-\eqref{eq:mna2} are
\begin{align}
    \mathbf{A}_{\textrm{C}}=\begin{bmatrix}
		 1 & 0\\
		-1 & 1
	\end{bmatrix},
    \mathbf{A}_{\textrm{R}}=\begin{bmatrix}
		0\\
		1
	\end{bmatrix}
    \text{ and }
    \mathbf{A}_{\textrm{I}}=\begin{bmatrix}
		1\\
		0
	\end{bmatrix}
 \end{align}
 and the current $I=\SI{1}{\ampere}$. However, this current excitation does not matter for the following stability analysis. The first stabilization (i) yields
\begin{align}
	\begin{bmatrix}
		R^{-1}+j\omega C & -j\sqrt{\omega}C\\
		-j\sqrt{\omega}C & j2C
	\end{bmatrix}
    \!
	\begin{bmatrix}
		\phi_1\\
		\xi_2
	\end{bmatrix}
	=   
	\begin{bmatrix}
		-I\\
		0
	\end{bmatrix}
 	\nonumber
\end{align}
with $\xi_2=\sqrt{\omega}\phi_2$. The corresponding condition number is given in the limit $\omega\to0$ by
\begin{align}
	\kappa_{(i)}&=\frac{1}{2RC}+\mathcal{O}(\sqrt{\omega})
\end{align}
where we used the assumption of $R^{-1}>2\omega C$. The condition number can be improved when including the parameters of the lumped elements in the scaling, e.g, $a_1 = b_1 =1/\sqrt{R^{-1}+j\omega C}$ and $a_2=b_2=1/\sqrt{j\omega2C}$. This corresponds to variant (iii) and yields
\begin{align}
	\!\!\begin{bmatrix}
		1 
        &
        \frac{1+j}{-2}
        \sqrt{\frac{\omega C}{R^{-1}+j\omega C}}
        \\
        \frac{1+j}{-2}
        \sqrt{\frac{\omega C}{R^{-1}+j\omega C}}
        &
        1
	\end{bmatrix}
    \!
	\begin{bmatrix}
		\xi_1\\
		\xi_2
	\end{bmatrix}
	=   
	\begin{bmatrix}
		\frac{-I}{\sqrt{R^{-1}+j\omega C}}\\
		0
	\end{bmatrix}
	\nonumber
\end{align}
which scales both unknowns $\xi_k=\varphi_k/b_k$. The corresponding condition number is in the limit $\omega\to0$ even optimal:
\begin{align}
	\kappa_{(iii)}&=1+\mathcal{O}(\sqrt{\omega}).
\end{align}
Note, in both cases the potential $\phi_2$ cannot be obtained from the unknown $\xi_2$ if $\omega=0$. 

\autoref{fig:condition_circuit} visualizes the condition numbers for two capacitors of $C=\SI{1e12}{\farad}$ and one resistor $R=\SI{1}{\ohm}$. 
The data is derived semi-analytically using MATHEMATICA\textsuperscript{\textregistered}~\cite{Mathematica_2020aa}. 
Thus, the computational cost is negligible and a very large frequency range is computable, i.e., $f=\SI{1e-20}{\hertz}$ to $\SI{1e40}{\hertz}$. 
The plot shows that the original formulation breaks down for frequencies below $\SI{1e10}{\hertz}$, while formulations (i) and (iii) remain stable down to \SI{0}{\hertz}. 
Formulation (iii) has the same qualitative behavior for low frequencies as (i) but reduces the condition number by $\num{5e11}$, i.e., the quotient $1/(2RC)$, as expected. 
We observe that formulation (i) destabilizes for high frequencies above $\SI{1e10}{\hertz}$. This is also to be expected, since the conductive contribution becomes negligible and the problem is essentially capacitive. Similar behavior is also known from stabilizations of full Maxwell formulations, see, e.g., Figure~5 in \cite{Eller_2017aa}.

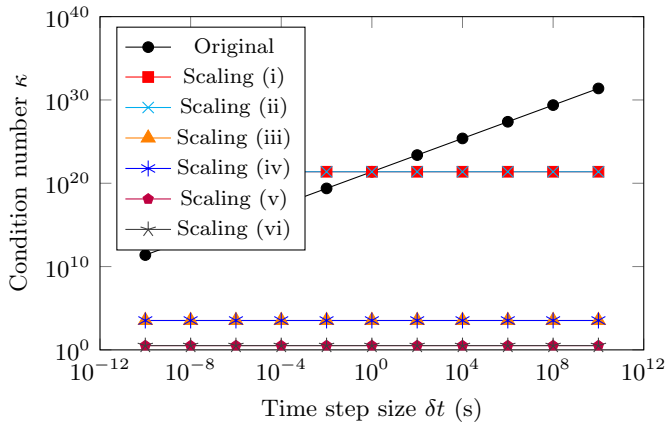
\begin{figure}
	\centering
	\begin{tikzpicture}
    \begin{loglogaxis}[
	width=0.99\columnwidth,
	height=6cm,
    xlabel={Time step size $\delta t$ (\si{s})},
    ylabel={Condition number $\kappa$},
    ytick={1e0, 1e10, 1e20, 1e30, 1e40},
    ymax=1e40,
    ymin=1,
    tick label style={font=\small},
    legend pos={north west},
	legend style={font=\footnotesize}, 
    label style={font=\small},
    every x tick scale label/.style={at={(1,0)},
    anchor=north,yshift=-5pt,inner sep=0pt}]
        \addplot [black, mark=*] table [x index=0, y index=1, col sep=comma]{data/conditions_tdei_h_v1-6.csv};
        \addplot [red, mark=square*] table [x index=0, y index=2, col sep=comma]{data/conditions_tdei_h_v1-6.csv};
        \addplot [cyan, mark=x, mark size=3pt] table [x index=0, y index=3, col sep=comma]{data/conditions_tdei_h_v1-6.csv};
        \addplot [orange, mark=triangle*, mark size=3pt] table [x index=0, y index=4, col sep=comma]{data/conditions_tdei_h_v1-6.csv};
        \addplot [blue, mark=asterisk, mark size=3pt] table [x index=0, y index=5, col sep=comma]{data/conditions_tdei_h_v1-6.csv};
        \addplot [purple, mark=pentagon*] table [x index=0, y index=6, col sep=comma]{data/conditions_tdei_h_v1-6.csv};
        \addplot [darkgray, mark=star, mark size=3pt] table [x index=0, y index=7, col sep=comma]{data/conditions_tdei_h_v1-6.csv};
        \legend{
            Original\\ 
            Scaling (i)\\ 
            Scaling (ii)\\ 
            Scaling (iii) \\ 
            Scaling (iv) \\
            Scaling (v)\\
            Scaling (vi)\\
        }
    \end{loglogaxis}
	\end{tikzpicture}
	\caption{Condition number of the EQS example in dependence of time step size. The breakdown for $\omega\to0$ translates her to $\delta t\to\infty$.}
	\label{fig:condition_eqs}
\end{figure}

\subsection{Electroquasistatic Toy Example}
As a numerical toy example we investigate a layered parallel plate capacitor in the time domain whose plates are connected by a conductor (\autoref{fig:anordnung}). This benchmark was originally proposed in \cite{Ostrowski_2021aa} and is constructed such that it yields a homogeneous electric displacement field in the whole domain. The computational domain $\Omega$ is a cube of side length $d=\SI{22}{\centi\metre}$, see \autoref{fig:anordnung}. The domain is subdivided into $\Omega_{\mathrm{i}}$ (blue region) and $\Omega_{\mathrm{o}}$ (remaining region). $\Omega_{\mathrm{i}}$ extends along the entire length in $y$- and $z$-directions. The red region extends in the $x$-direction and represents a conductor. The other region is non-conducting. The $\Omega_{\mathrm{i}}$ region has permittivity $\varepsilon_{\mathrm{i}} = \varepsilon_0$ and conductivity $\sigma_{\mathrm{i}} = \SI{2.98e7}{\siemens\per\meter}$. The region $\Omega_{\mathrm{o}}$ has permittivity $\varepsilon_{\mathrm{o}} = 2\varepsilon_0$ and conductivity $\sigma_{\mathrm{o}} = 2\sigma_{\mathrm{i}} = \SI{5,96e7}{\siemens\per\meter}$. For the geometry parameters, $d_{\mathrm{o}} = \SI{10}{\centi\meter}$ and $d_{\mathrm{i}} = \SI{2}{\centi\meter}$.
At the left and right boundaries of the box, Dirichlet  conditions are set for the electric potential $\varphi$. The boundary conditions are $\varphi_{\mathrm{bc},1}=\SI{0}{\volt}$ and $\varphi_{\mathrm{bc},2} = (\SI{1}{V})\sin(\omega t)$ with angular frequency $\omega = 2\pi f$ and $f = \SI{50}{\hertz}$. Homogeneous Neumann conditions are present at the remaining boundaries. Moreover, no impressed electric charges or current densities are given. The excitation is realized solely by the boundary conditions. 

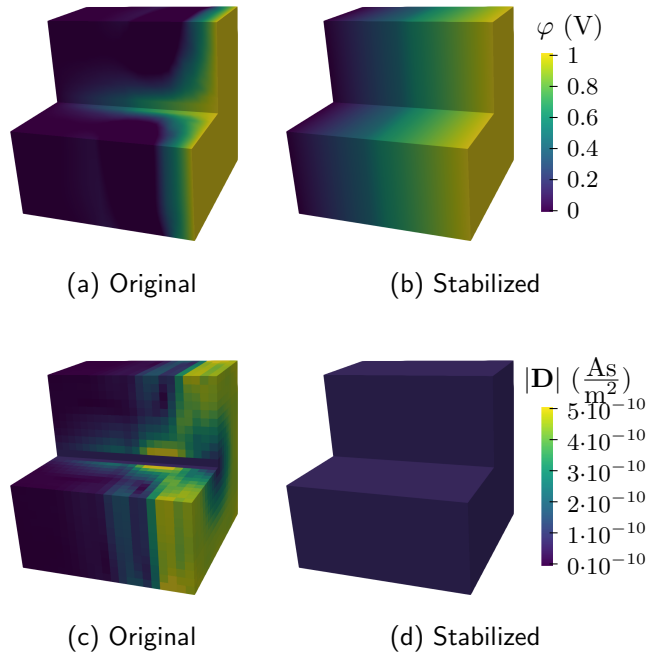
\begin{figure}
    \centering
    \subfloat[Original]{%
        \begin{tikzpicture}\node at (0,0) {
                \includegraphics[height=3.5cm,trim={4cm 0 5.7cm 0},clip]{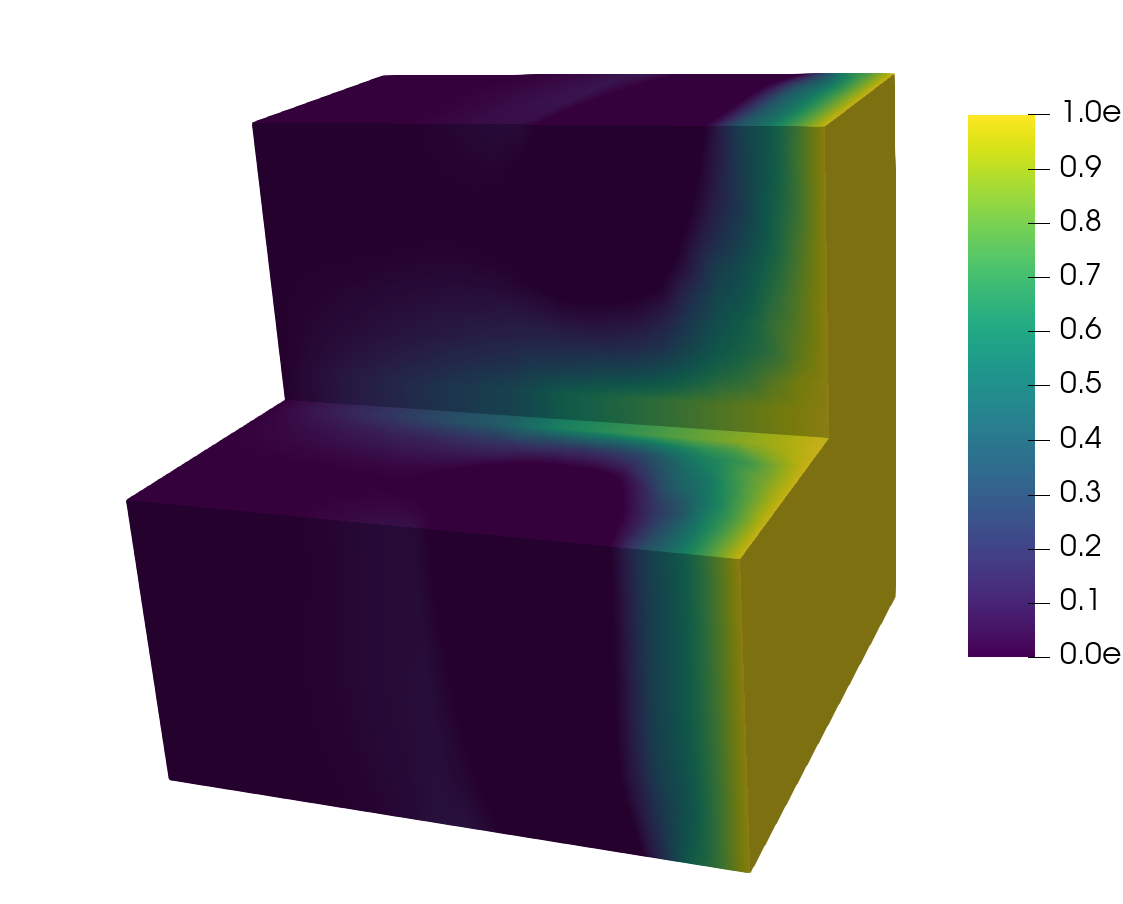}};
            \end{tikzpicture}}
    \subfloat[Stabilized]{
        \begin{tikzpicture}\node at (0,0) {
            \includegraphics[height=3.5cm,trim={4cm 0 3.3cm 0},clip]{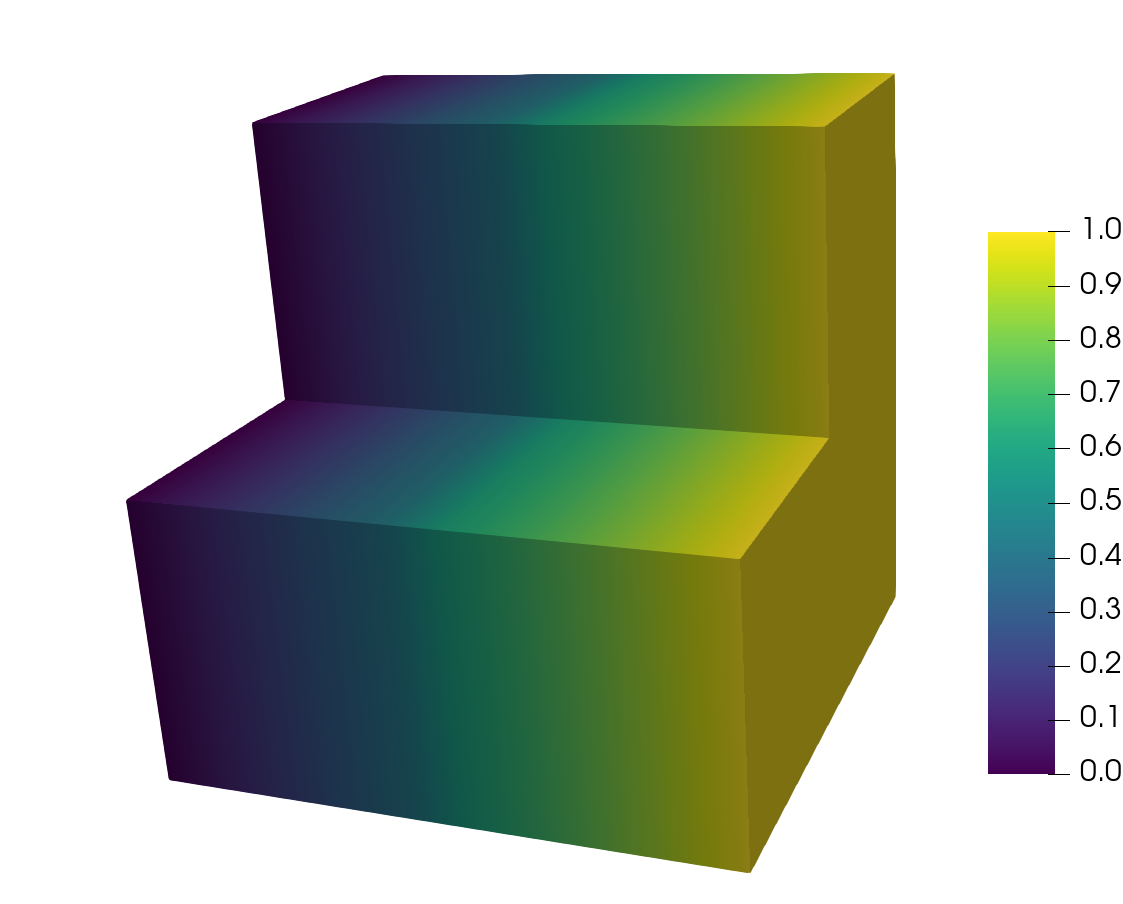}};
            \foreach \i in {0,0.2,...,1}{  
                \draw (1.73,2.06*\i-1.23)--++(3pt,0) node [anchor=west] {\small\pgfmathprintnumber{\i}};
            }
            \phantom{\draw (1.73,2.06*5/5-1.23)--++(3pt,0) node [anchor=west] {\small\pgfmathprintnumber{5}$\cdot 10^{-10}$};}
            \node at (2.0, 1.2) {$\varphi$ (V)};
        \end{tikzpicture}
    }\\
    \subfloat[Original]{%
        \begin{tikzpicture}\node at (0,0) {
                \includegraphics[height=3.5cm,trim={4cm 0 5.7cm 0},clip]{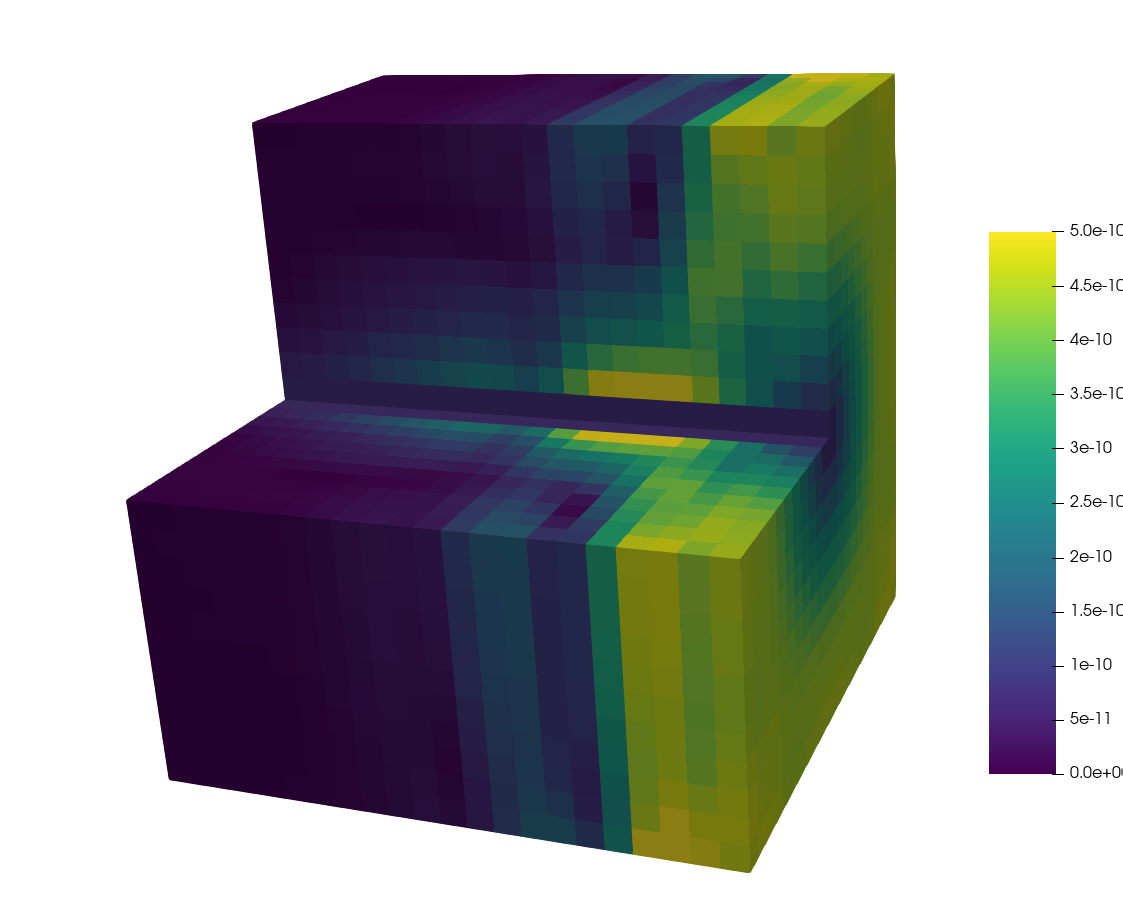}};
            \end{tikzpicture}}
    \subfloat[Stabilized]{
        \begin{tikzpicture}\node at (0,0) {
            \includegraphics[height=3.5cm,trim={4cm 0 3.3cm 0},clip]{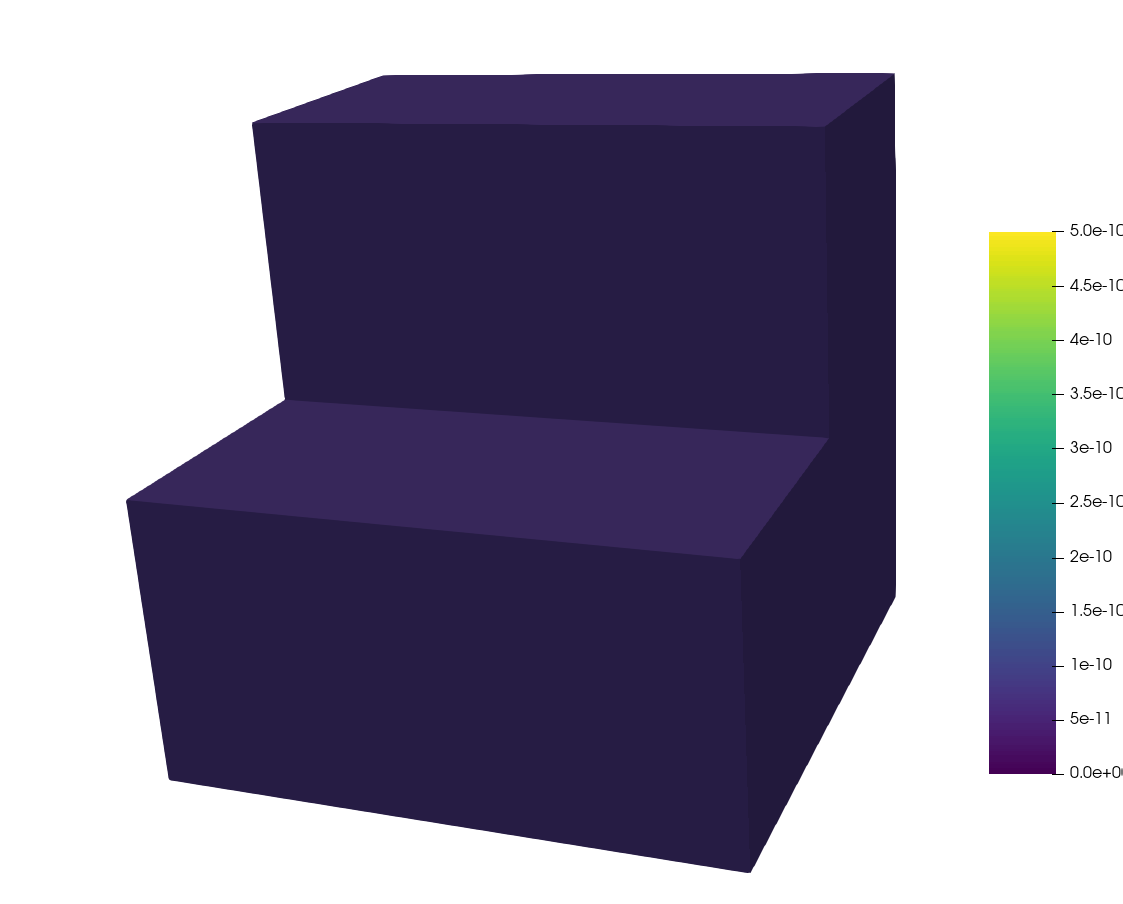}};
            \foreach \i in {0,1,...,5}{  
                \draw (1.73,2.06*\i/5-1.23)--++(3pt,0) node [anchor=west] {\small\pgfmathprintnumber{\i}$\cdot 10^{-10}$};
            }
            \node at (2.1, 1.2) {$\left|\mathbf{D}\right|$ ($\frac{\si{As}}{\si{m^2}}$)};
        \end{tikzpicture}
    }
    \vspace{0.5em}
    \caption{Visualization of the electric scalar potential $\varphi$ and displacement field $\left|\mathbf{D}\right|$ at peak voltage for the configuration given in \autoref{fig:anordnung}. The correct analytical solution is a constant displacement field $\left|\mathbf{D}\right|\approx\SI{7.38e-11}{As\per m^2}$. Shown are results for the original (left, instable) and (iv)-th formulation (right, stable). The time step was $\delta t=\SI{1}{\milli\second}$ and the iterative solver used was the  MATLAB\textsuperscript{\textregistered} function `bicgstab' without preconditioner. The original problem took 565 iterations and variant (iv) 232 iterations to compute for a tolerance of $\num{1e-15}$.}
    \label{fig:field_phi}
\end{figure}

The model is discretized by $N=11109$ degrees of freedom defined on hexahedral elements of lowest order.
From an engineering perspective, an adequate time step size for the given right hand side would be $\delta t=\SI{2}{\milli\second}$ such that it resolves each period of the sinusoidal excitation adequately. 
We investigate the condition of the system matrix that results from one step of the implicit Euler method \eqref{eq:euler} for varying formulations (i-vi) and time step sizes $\delta t=\SI{1e-10}{\second},\ldots,\SI{1e10}{\second}$. 
The condition number is estimated using the $\infty$-norm based on the MATLAB\textsuperscript{\textregistered} function `condest'~\cite{Mathworks_2020aa}. \autoref{fig:condition_eqs} shows the results. 
For illustration purposes, \autoref{fig:field_phi} depicts the resulting field of the original and a stabilized formulation. Note, that only the stabilized formulation gives the correct behavior, i.e., constant displacement field.

All stabilizations improve the condition number for $\delta t\to\infty$ (low frequency translates to large time step size). 
The variants (iii,iv) involving material coefficients significantly reduce the condition number ($\sim10^{18}$) and perform well over the entire frequency range. However, in contrast to the circuit example, we are not dealing with scalars but matrices. Therefore, the condition number remains around $\kappa\approx 10^{3}$. Finally, the variants (v,vi)
use the MATLAB\textsuperscript{\textregistered} function `ilu' and bring down the condition number to approximately $\kappa\approx 3$. 

\subsection{High-Voltage Bushing}
\begin{figure}[tb]
    \centering
    \subfloat[Geometry]{\includegraphics[trim=4cm 1cm 4cm 0.6cm, clip, width=2.5cm]{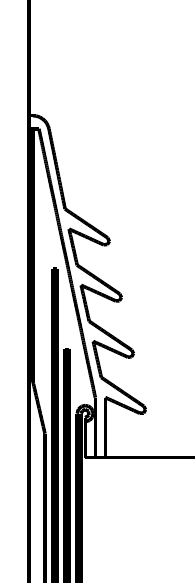}}
    \hfill
    \subfloat[Mesh]{\includegraphics[trim=4cm 1cm 4cm 0.6cm, clip, width=2.5cm]{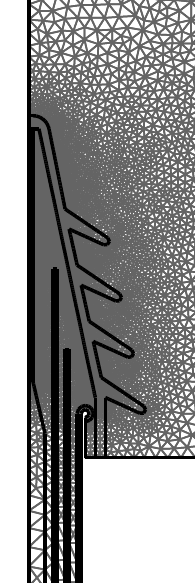}}
    \hfill
    \subfloat[Instability]{\includegraphics[trim=4cm 1cm 4cm 0.6cm, clip, width=2.5cm]{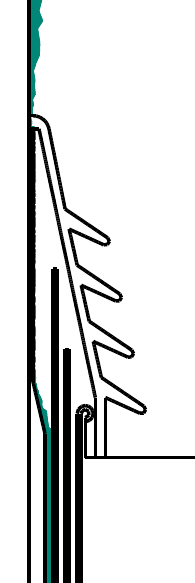}}
    \vspace{0.5em}
    \caption{Geometry, mesh and unstable solution for the high-voltage bushing.}
    \label{fig:roth2_bushing}
\end{figure}

An oil-filled condenser-type bushing is used here as an example of industrial relevance \cite{Kuffel_2000aa,Monga_2006aa} (\autoref{fig:roth2_bushing}). The electric stress at the triple-junction point is diminished by a ground electrode reaching into the bushing and by two additional, concentric, metallic cylinders mounted between the central high-voltage electrode and the outer grounded electrode. The oil has a conductivity $\sigma_{\rm oil}=\SI{1.0}{\pico\siemens\per\meter}$ and a permittivity $\varepsilon_{\rm oil}=2.2\varepsilon_0$. The porcelain housing has a permittivity $\varepsilon_{\rm prc}=6.5\varepsilon_0$. The electroquasistatic model is simulated in frequency domain using an in-house low-order axisymmetric 2D finite element solver considering 7248 degrees of freedom. All metallic parts are considered as perfectly conducting and modeled by floating-potential conditions.

\begin{figure}[tb]
    \centering
    \subfloat[\SI{0.5}{\hertz}]{\includegraphics[trim=4cm 1cm 4cm 0.6cm, clip, width=2.5cm]{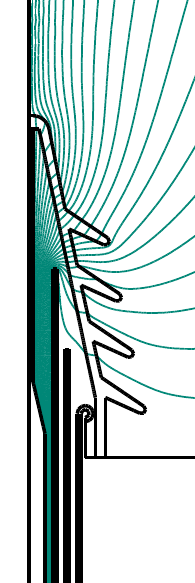}}
    \hfill
    \subfloat[\SI{2}{\hertz}]{\includegraphics[trim=4cm 1cm 4cm 0.6cm, clip, width=2.5cm]{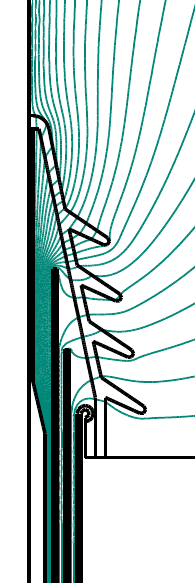}}
    \hfill
    \subfloat[\SI{50}{\hertz}]{\includegraphics[trim=4cm 1cm 4cm 0.6cm, clip, width=2.5cm]{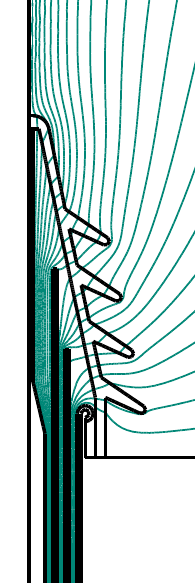}}
    \vspace{0.5em}
    \caption{Stably computed electroquasistatic field in the high-voltage bushing for three operating frequencies computed using variant (iii)}.
    \label{fig:roth2_equipots}
\end{figure}

\autoref{fig:roth2_bushing}(c) illustrates the unstable behavior due the low-frequency breakdown depicting a wrong solution computed using a sparse direct solver applied to the original problem formulation at \SI{0}{\hertz}. For comparison, the correct behavior is shown in \autoref{fig:roth2_equipots} for three different frequencies.

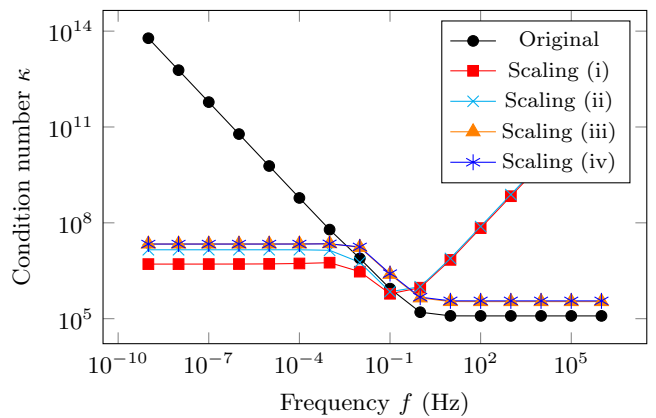
\begin{figure}
	\centering
	\begin{tikzpicture}
    \begin{loglogaxis}[
	width=0.99\columnwidth,
	height=6cm,
    xlabel={Frequency $f$ (Hz)},
    ylabel={Condition number $\kappa$},
    tick label style={font=\small},
    legend pos={north east},
	legend style={font=\footnotesize}, 
    label style={font=\small},
    every x tick scale label/.style={at={(1,0)},
    anchor=north,yshift=-5pt,inner sep=0pt}]
        \addplot [black, mark=*] table [x index=0, y index=1, col sep=comma]{data/bushing_cnff.csv};
        \addplot [red, mark=square*] table [x index=0, y index=2, col sep=comma]{data//bushing_cnff.csv};
        \addplot [cyan, mark=x, mark size=3pt] table [x index=0, y index=3, col sep=comma]{data//bushing_cnff.csv};
        \addplot [orange, mark=triangle*, mark size=3pt] table [x index=0, y index=4, col sep=comma]{data//bushing_cnff.csv};
        \addplot [blue, mark=asterisk, mark size=3pt] table [x index=0, y index=5, col sep=comma]{data//bushing_cnff.csv};
        \legend{
            Original\\ 
            Scaling (i)\\ 
            Scaling (ii)\\ 
            Scaling (iii) \\ 
            Scaling (iv) \\
        }
    \end{loglogaxis}
	\end{tikzpicture}
	\caption{Condition number of the high-voltage bushing example as a function of frequency.}
	\label{fig:roth2_condition}
\end{figure}

\autoref{fig:roth2_condition} shows the condition numbers for the first four stabilization variants over frequency. The condition number was again estimated using the $\infty$-norm based on the MATLAB\textsuperscript{\textregistered} function `condest',~\cite{Mathworks_2020aa}. The results for this more realistic configuration remain similar as for the previous studies and confirm the theory: all variants significantly improve the low-frequency behavior down to \SI{0}{\hertz}. As before, the first two variants (i) and (ii) can lead to problems for high frequencies which are not relevant in the electroquasistatic regime. Variants (iii) and (iv) lead to excellent results for all considered frequencies. The Jacobi-type variants (v)-(vi) are not included since it cannot be guaranteed in general that all conducting parts touch Dirichlet boundaries.


\subsection{Discussion}
Let us compare the variants from a practical point of view: If a direct solver for sparse symmetric linear systems is used, then formulations (i) or (iii) are well suited. The latter gives better results but requires knowledge on material data which may be inconvenient to implement {-- or one uses a Jacobi-type implementation \eqref{eq:jacobi1}-\eqref{eq:jacobi2}}. If the direct solver does not exploit symmetry, then (ii) and (iv) are good choices. They have the additional benefit that no rescaling after the solution process is necessary. Note, that some sparse direct solvers, in particular UMFPACK \cite{Davis_2004aa}, apply their own scaling based on heuristics, which may diminish the effect of manual scaling (if $\omega>0$) but it cannot harm either since the numerical effort is negligible. 

Finally, if an iterative solver is used, it depends on the available preconditioners. For example the variant (vi) is a computationally cheap and appropriate choice, e.g., for a frequency-sweep in the low frequency regime.

\section{Conclusion}
We have discussed the low-frequency instability of electroquasistatic problems that is less well-known than the one of full-wave formulations. 
We proposed several simple scalings of the system matrix for time and frequency domain to circumvent the breakdown. All approaches are effective and almost trivial to implement. The numerical examples confirm the necessity of low-frequency stabilization for dielectric circuit and field problems.

\section*{Acknowledgments}
The authors thank Markus Clemens for many fruitful discussion on electroquasistatic and Darwin field formulations. Funding of the Graduate School CE at TU~Darmstadt and DFG (CRC TRR 361 and SCHO 1562/6-1) is acknowledged.

\renewcommand*{\bibfont}{\footnotesize}
\printbibliography

@BOOK{Haus_1989aa,
  AUTHOR = {Haus, Hermann A. and Melcher, James R.},
  PUBLISHER = {Prentice-Hall},
  URL = {http://web.mit.edu/6.013_book/www/},
  DATE = {1989},
  ISBN = {978-0-13-249020-7},
  LANGID = {english},
  TITLE = {Electromagnetic Fields and Energy},
}

@ARTICLE{Dirks_1996aa,
  AUTHOR = {Dirks, Heinz K.},
  DATE = {1996},
  DOI = {10.1007/BF01232924},
  ISSN = {1432-0487},
  JOURNALTITLE = {Electr. Eng.},
  LANGID = {english},
  NUMBER = {2},
  PAGES = {145--155},
  TITLE = {Quasi-Stationary Fields for Microelectronic Applications},
  VOLUME = {79},
}

@ARTICLE{van-Rienen_1996aa,
  AUTHOR = {van Rienen, Ursula and Clemens, Markus and Weiland, Thomas},
  DATE = {1996-05-01},
  DOI = {10.1109/20.497366},
  ISSN = {0018-9464},
  JOURNALTITLE = {{IEEE} Trans. Magn.},
  NUMBER = {3},
  PAGES = {816--819},
  TITLE = {Simulation of low-frequency fields on high-voltage insulators with light contaminations},
  VOLUME = {32},
}

@ARTICLE{Clemens_1998aa,
  AUTHOR = {Clemens, Markus and Weiland, Thomas and van Rienen, Ursula},
  DATE = {1998-09-01},
  DOI = {10.1109/20.717784},
  ISSN = {0018-9464},
  JOURNALTITLE = {{IEEE} Trans. Magn.},
  NUMBER = {5},
  PAGES = {3335--3338},
  TITLE = {Comparison of {Krylov}-type methods for complex linear systems applied to high-voltage problems},
  VOLUME = {34},
}

@ARTICLE{Monga_2006aa,
  AUTHOR = {Monga, S. and Gorur, R. and Hansen, P. and Massey, W.},
  DATE = {2006-12},
  DOI = {10.1109/tdei.2006.258193},
  ISSN = {1070-9878},
  JOURNALTITLE = {{IEEE} Trans. Dielectr. Electr. Insul.},
  NUMBER = {6},
  PAGES = {1217--1224},
  TITLE = {Design optimization of high voltage bushing using electric field computations},
  VOLUME = {13},
}

@ARTICLE{Weida_2009aa,
  AUTHOR = {Weida, Daniel and Steinmetz, Thorsten and Clemens, Markus},
  DATE = {2009-03},
  DOI = {10.1109/TMAG.2009.2012492},
  ISSN = {0018-9464},
  JOURNALTITLE = {{IEEE} Trans. Magn.},
  LANGID = {english},
  NUMBER = {3},
  PAGES = {980--983},
  TITLE = {Electro-Quasistatic High Voltage Field Simulations of Large Scale Insulator Structures Including {2-D} Models for Nonlinear Field-Grading Material Layers},
  VOLUME = {45},
}

@ARTICLE{Christen_2010aa,
  AUTHOR = {Christen, Thomas and Donzel, Lise and Greuter, Felix},
  DATE = {2010-11-01},
  DOI = {10.1109/MEI.2010.5599979},
  ISSN = {1056-9170},
  JOURNALTITLE = {{IEEE} Electr. Insul. Mag.},
  NUMBER = {6},
  PAGES = {47--59},
  TITLE = {Nonlinear resistive electric field grading, Part 1: Theory and Simulation},
  VOLUME = {26},
}

@INPROCEEDINGS{Zhang_2010aa,
  AUTHOR = {Zhang, Chao and Kester, Jeffrey J. and Daley, Charles W. and Rigby, Stephen J.},
  BOOKTITLE = {2010 Annual Report Conference on Electrical Insulation and Dielectric Phenomena ({CEIDP})},
  DATE = {2010-10},
  DOI = {10.1109/CEIDP.2010.5723955},
  LANGID = {english},
  PAGES = {1--4},
  TITLE = {Electric field analysis of high voltage apparatus using finite element method},
}

@ARTICLE{Clemens_2022aa,
  AUTHOR = {Clemens, Markus and Henkel, Marvin-Lucas and Kasolis, Fotios and Günther, Michael and De Gersem, Herbert and Schöps, Sebastian},
  URL = {https://www.compumag.org/wp/newsletter/},
  DATE = {2022},
  EPRINT = {2204.06286},
  EPRINTTYPE = {arxiv},
  ISSN = {1026-0854},
  JOURNALTITLE = {{ICS} Newslett.},
  LANGID = {english},
  NUMBER = {1},
  PAGES = {3--9},
  TITLE = {Quasistatic Darwin Model Field Formulations in Time Domain},
  VOLUME = {29},
}

@ARTICLE{Hiptmair_2008aa,
  AUTHOR = {Hiptmair, R. and Kramer, F. and Ostrowski, J.},
  DATE = {2008-06},
  DOI = {10.1109/tmag.2007.915991},
  ISSN = {0018-9464},
  JOURNALTITLE = {{IEEE} Trans. Magn.},
  NUMBER = {6},
  PAGES = {682--685},
  TITLE = {A Robust {Maxwell} Formulation for All Frequencies},
  VOLUME = {44},
}

@ARTICLE{Jochum_2015aa,
  AUTHOR = {Jochum, Martin Thomas and Farle, Ortwin and Dyczij-Edlinger, Romanus},
  DATE = {2015-03},
  DOI = {10.1109/TMAG.2014.2360080},
  ISSN = {0018-9464},
  JOURNALTITLE = {{IEEE} Trans. Magn.},
  NUMBER = {3},
  PAGES = {7402304},
  TITLE = {A new low-frequency stable potential formulation for the finite-element simulation of electromagnetic fields},
  VOLUME = {51},
}

@ARTICLE{Eller_2017aa,
  AUTHOR = {Eller, Martin and Reitzinger, Stefan and Schöps, Sebastian and Zaglmayr, Sabine},
  DATE = {2017-08-24},
  DOI = {10.1137/16M1077817},
  ISSN = {1064-8275},
  JOURNALTITLE = {{SIAM} J. Sci. Comput.},
  LANGID = {english},
  NUMBER = {4},
  PAGES = {B703--B731},
  TITLE = {A Symmetric Low-Frequency Stable Broadband {Maxwell} Formulation for Industrial Applications},
  VOLUME = {39},
}

@ARTICLE{Stysch_2022aa,
  AUTHOR = {Stysch, Jonathan and Klaedtke, Andreas and Gersem, Herbert De},
  DATE = {2022-06},
  DOI = {10.1109/temc.2021.3134323},
  ISSN = {0018-9375},
  JOURNALTITLE = {{IEEE} Trans. {EMC}},
  NUMBER = {3},
  PAGES = {750--759},
  TITLE = {Low-Frequency Stabilization for {FEM} Impedance Computation},
  VOLUME = {64},
}

@ARTICLE{Zhao_2019aa,
  AUTHOR = {Zhao, Yanpu and Tang, Zuqi},
  DATE = {2019-06},
  DOI = {10.1109/tmag.2019.2896647},
  ISSN = {0018-9464},
  JOURNALTITLE = {{IEEE} Trans. Magn.},
  NUMBER = {6},
  PAGES = {1--4},
  TITLE = {A Symmetric Field-Circuit Coupled Formulation for 3-D Transient Full-Wave {Maxwell} Problems},
  VOLUME = {55},
}

@ARTICLE{Ostrowski_2021aa,
  AUTHOR = {Ostrowski, Jörg and Hiptmair, Ralf},
  DATE = {2021-01},
  DOI = {10.1137/20m1356300},
  ISSN = {1064-8275},
  JOURNALTITLE = {{SIAM} J. Sci. Comput.},
  NUMBER = {4},
  PAGES = {B1008--B1028},
  TITLE = {Frequency-Stable Full {Maxwell} in Electro-quasistatic Gauge},
  VOLUME = {43},
}

@INPROCEEDINGS{Kasolis_2021aa,
  AUTHOR = {Kasolis, Fotios and Henkel, Marvin-Lucas and Clemens, Markus},
  EDITOR = {Ehrhard, Matthias},
  LOCATION = {Wuppertal},
  ORGANIZATION = {European Consortium for Mathematics in Industry},
  BOOKTITLE = {21st European Conference on Mathematics for Industry ({ECMI} 2021)},
  DATE = {2021-04-13},
  ENTRYSUBTYPE = {talk},
  LANGID = {english},
  TITLE = {Low-Frequency Stable Electro-Quasistatic Field Formulations Based on Continuous Extensions},
}

@BOOK{Saad_2000aa,
  AUTHOR = {Saad, Yousef},
  LOCATION = {Boston, MA, USA},
  PUBLISHER = {Society for Industrial {and} Applied Mathematics},
  URL = {http://www-users.cs.umn.edu/~saad/books.html},
  DATE = {2003},
  EDITION = {2},
  ISBN = {978-0-89871-534-7},
  LANGID = {english},
  TITLE = {Iterative Methods for Sparse Linear Systems},
}

@ARTICLE{Ho_1975aa,
  AUTHOR = {Ho, Chung-Wen and Ruehli, Albert E. and Brennan, Pierce A.},
  DATE = {1975-06},
  DOI = {10.1109/TCS.1975.1084079},
  JOURNALTITLE = {{IEEE} Trans. Circ. Syst.},
  LANGID = {english},
  NUMBER = {6},
  PAGES = {504--509},
  TITLE = {The Modified Nodal Approach to Network Analysis},
  VOLUME = {22},
}

@BOOK{Monk_2003aa,
  AUTHOR = {Monk, Peter},
  LOCATION = {Oxford},
  PUBLISHER = {Oxford University Press},
  DATE = {2003},
  LANGID = {english},
  TITLE = {Finite Element Methods for {Maxwell}'s Equations},
}

@BOOK{Brenner_2008aa,
  AUTHOR = {Brenner, Susanne C. and Scott, Larkin Ridgway},
  LOCATION = {New York},
  PUBLISHER = {Springer},
  DATE = {2008},
  EDITION = {3. ed.},
  ISBN = {9780387759333},
  SERIES = {Texts in applied mathematics},
  TITLE = {The mathematical theory of finite element methods},
  VOLUME = {15},
}

@ARTICLE{Dular_1998aa,
  AUTHOR = {Dular, Patrick and Legros, Willy and Nicolet, André},
  DATE = {1998-09},
  ISSN = {0018-9464},
  JOURNALTITLE = {{IEEE} Trans. Magn.},
  LANGID = {english},
  NUMBER = {5},
  PAGES = {3018--3021},
  TITLE = {Coupling of local and global quantities in various finite element formulations its application to electrostatics, magnetostatics and magnetodynamics},
  VOLUME = {34},
}

@ARTICLE{De-Gersem_2003ac,
  AUTHOR = {De Gersem, Herbert and Belmans, Ronnie J. M. and Hameyer, Kay},
  DATE = {2003},
  ISSN = {0332-1649},
  JOURNALTITLE = {{COMPEL}},
  LANGID = {english},
  NUMBER = {1},
  PAGES = {20--29},
  TITLE = {Floating potential constraints and field-circuit couplings for electrostatic and electrokinetic finite element models},
  VOLUME = {22},
}

@BOOK{Hairer_1996aa,
  AUTHOR = {Hairer, Ernst and Nørsett, Syvert P. and Wanner, Gerhard},
  LOCATION = {Berlin, Germany},
  PUBLISHER = {Springer},
  DATE = {2002},
  EDITION = {2},
  ISBN = {978-3-642-05220-0},
  LANGID = {english},
  SERIES = {Springer Series in Computational Mathematics},
  TITLE = {Solving Ordinary Differential Equations {II}: Stiff and Differential-Algebraic Problems},
}

@MANUAL{Mathematica_2020aa,
  AUTHOR = {Wolfram},
  URL = {https://www.wolfram.com/mathematica/},
  DATE = {2020},
  TITLE = {Mathematica},
  URLDATE = {2020-01-02},
}

@MANUAL{Mathworks_2020aa,
  AUTHOR = {Mathworks},
  DATE = {2020},
  LANGID = {english},
  NOTE = {9.8.0 (R2020a)},
  TITLE = {{MATLAB} Getting Started Guide},
}

@BOOK{Kuffel_2000aa,
  AUTHOR = {Kuffel, E. and Zaengl, W. S. and Kuffel, D.},
  PUBLISHER = {Newnes},
  DATE = {2000},
  EDITION = {2nd},
  LANGID = {english},
  TITLE = {High Voltage Engineering - Voltage},
}

@ARTICLE{Davis_2004aa,
  AUTHOR = {Davis, Timothy A.},
  DATE = {2004-06},
  DOI = {10.1145/992200.992206},
  ISSN = {0098-3500},
  JOURNALTITLE = {{ACM} Trans. Math. Software},
  NUMBER = {2},
  PAGES = {196--199},
  TITLE = {Algorithm 832: {UMFPACK} V4.3 -- an unsymmetric-pattern multifrontal method},
  VOLUME = {30},
}
\end{document}